# Switchable Weyl nodes in topological Kagome ferromagnet $Fe_3Sn_2$

M. Yao[1*], H. Lee[2*], N. Xu[1§], Y. Wang[3,4], J. Ma[1], O. V. Yazyev[2], Y. Xiong[3,5], M. Shi[1], G. Aeppli[1,2,6], and Y. Soh[1]

[1]Paul Scherrer Institut, Forschungsstrasse 111, 5232 Villigen PSI, Switzerland

[2]Institute of Physics, Ecole Polytechnique Fédérale de Lausanne (EPFL), CH-1015 Lausanne, Switzerland

[3]Anhui Province Key Laboratory of Condensed Matter Physics at Extreme Conditions, High Magnetic Field Laboratory, Chinese Academy of Sciences, Hefei, Anhui 230031, China

[4]University of Science and Technology of China, Hefei, Anhui 230026, China

[5]Collaborative Innovation Center of Advanced Microstructures, Nanjing, 210093, China

[6]Laboratory for Solid State Physics, ETH Zurich, Zurich, CH-8093, Switzerland

[*] Both authors made equal contribution to the project

[§]Present address: The Institute of Advanced Studies, Wuhan University, Wuhan 430072, China

**Abstract:** The control of topological quantum materials is the prerequisite for novel devices exploiting these materials. Here we propose that the room temperature ferromagnet $Fe_3Sn_2$, whose fundamental building blocks are Kagome bilayers of iron, hosts Weyl nodes at the Fermi level which can move in momentum space depending on the direction of the magnetization, itself readily controlled either by modest external fields or temperature. The proposal is derived from density functional calculations, including a mean field treatment of Hubbard repulsion U, which have been validated by comparison with angle-resolved photoemission data. Ferromagnetism with magnetization along certain directions is shown to lift the Weyl degeneracies, while at the same time inducing texture in the quasiparticle spin polarizations mapped in reciprocal space. In particular, the polarization is attenuated and then rotated from parallel to perpendicular to antiparallel to the magnetization as Weyl points derived from crossing of majority and minority spin bands are traversed.

Topological materials are interesting due to electrical properties which manifest quantum effects robust against small perturbations [1-4]. The quantum Hall (QH) state is the oldest topological system discovered and relies on 2-dimensionality and broken time-reversal symmetry $\mathcal{T}$ [5-6]. Therefore, for a long time it was thought that 2-dimensionlity and broken $\mathcal{T}$ were essential for topological states. However, this perception was changed with the discovery of graphene, which is 2-dimensional and hosts Dirac points at high symmetry points protected by $\mathcal{T}$ and inversion symmetry $\mathcal{P}$ [7-9]. Graphene hosts massless 2-dimensional Dirac fermions which display unusual large electronic mean free paths and exhibit both the integer and fractional quantum Hall effects.

Following the discoveries about graphene was the discovery of Dirac points and associated massless Dirac fermions for the surface states of three-dimensional materials called topological insulators, where the surface is conducting while the bulk is insulating [10]. While ordinary insulators can also host conducting surface states due to, for example, band bending, what is special about topological insulators is that their surface states have well-defined chiral spin textures, are robust against backscattering, and represent the two-dimensional analogue of the quantum spin Hall (QSH) effect edge states. Although there is a large similarity between the QH system and the QSH system, what is fundamentally distinct is that QH relies on broken $\mathcal{T}$ symmetry whereas the QSH effect relies on $\mathcal{T}$ symmetry and spin-orbit coupling.

More recently, interest has expanded to topological materials that host Weyl nodes, which are accidental touching points of non-degenerate bands corresponding to momentum-space monopoles that are sources or sinks of Berry curvature [11-12]. Based on Kramers theorem, systems with both $\mathcal{T}$ and inversion symmetry $\mathcal{P}$ are doubly degenerate at all points in momentum space, therefore, to host Weyl nodes, the system needs to have either $\mathcal{T}$ or $\mathcal{P}$ symmetry broken. In addition, strong spin-orbit coupling seems to be necessary to lift the degeneracy of bands interpenetrating each other to leave only isolated points where the bands touch each other. Weyl semimetals are systems where the measured electronic properties are dominated by Weyl nodes which are near the Fermi level ($E_F$). While the initial search for Weyl semimetals was first focused on systems with broken $\mathcal{T}$ symmetry, failure in the search led to the shift of focus to systems with broken $\mathcal{P}$ symmetry.

In systems with broken $\mathcal{P}$ symmetry, several Weyl semimetals have been predicted by density-functional-theory (DFT) calculations and confirmed using angle-resolved photoemission spectroscopy (ARPES) [13-17]. However, until now the search for Weyl semimetals hosted in systems with broken $\mathcal{T}$ has been harder. Here we propose $Fe_3Sn_2$ as a ferromagnet with $\mathcal{P}$ inversion symmetry but broken $\mathcal{T}$ symmetry that hosts Weyl nodes *at* $E_F$. Recently $Mn_3Sn$, which is an antiferromagnet and has a Kagome structure as does $Fe_3Sn_2$, has been suggested to host Weyl fermions near $E_F$ [18]. There have been also reports that ferromagnetic $Co_3Sn_2S_2$, where the transition metal atoms also appear in Kagome lattices, hosts Weyl nodes around 60 meV above $E_F$ [19-20]. What is special about $Fe_3Sn_2$ is not only that it is ferromagnetic at room temperature with a Curie temperature $T_C \cong 657$ K [21], in contrast to $Co_3Sn_2S_2$ which is paramagnetic with $T_C \cong 177$ K, but also that the predicted Weyl nodes are at or within a few meV of $E_F$, making them highly relevant to transport at room temperature and below, and that we can move the Weyl nodes not only with an external magnetic field but also with temperature.

The kagome-bilayer ferromagnet $Fe_3Sn_2$ has drawn attention for several reasons. Firstly, it undergoes an unexplained spin reorientation around $T_f \sim 100$ K [21-22]: on warming through $T_f$, the magnetization rotates from parallel to perpendicular to the kagome planes. In addition, a large anomalous Hall effect (AHE) was discovered on $Fe_3Sn_2$, which was attributed to non-trivial ferromagnetic texture [23]. Theorists have proposed $Fe_3Sn_2$ as a model system to exhibit flat bands with strong Coulomb interactions and therefore to display the fractional quantum Hall effect even at room temperature [24]. More recently, $Fe_3Sn_2$ has been suggested to host magnetic Skyrmions as well as Dirac-like points in the electronic band structure [25-26]. Therefore, $Fe_3Sn_2$ is a system hosting a wealth of exotic magnetic and electronic properties suggesting further exploration.

To understand the fascinating magnetic and electronic properties of $Fe_3Sn_2$, it is essential to start with a detailed band structure which takes into account at least at the mean field level the correlations responsible for the ferromagnetism. We meet this need by performing a comprehensive DFT+U calculation of both the surface and bulk states, where density functional theory (DFT) is combined with a Hubbard U correction. The calculation is validated by comparison with two- and three-dimensional features in the electronic structure seen by angle-resolved photoemission spectroscopy (ARPES) for UV and soft X-ray light, respectively. In particular, it accounts for the quasi-2D Dirac-like cone seen previously [26]. What is new and more exciting is the prediction of features beyond the resolution of measurements performed hitherto, in particular Weyl nodes which are at or very close to $E_F$. The nodes move and indeed occur at level crossings which are avoided or not depending on the magnetization direction. Furthermore, when magnetization eliminates the nodes, the associated quasiparticles exhibit spin texture in reciprocal space such that their polarizations rotate perpendicular to the magnetization at avoided crossings.

Fe$_3$Sn$_2$ has the layered rhombohedral structure (space group R-3m) shown in Fig. 1a. The Fe ions form bilayers of kagome networks separated by Sn layers. Shorter and larger equilateral triangles build the kagome layers, as shown in Fig. 1b. The crystal structure preserves the $\mathcal{P}$ symmetry and 3-fold rotational symmetry. As the system is ferromagnetic, $\mathcal{T}$ is broken.

Figure 1c shows the bulk and surface Brillouin zone with high symmetry points indicated. In Fig. 1d we plot the bulk band structures of Fe$_3$Sn$_2$ for the different magnetization directions from first-principles calculations with spin-orbit coupling included. To treat the correlations between electrons in the Fe-$d$ orbitals, we add the simplified Hubbard $U$ correction to Fe-$d$ orbitals. By comparing the calculated spectra with ARPES data, a value of $U_{\text{eff}}=U-J=0.5$ eV is chosen. The choice of U=0.5 eV gives us Dirac-like features at $K$ and $K'$ near $E-E_F = -0.1$ eV, in agreement with the present (see more detail below) and previous experimental data. Experimentally, Fe$_3$Sn$_2$ favors in-plane magnetization at low temperature. Our total energy calculations confirm this with an energy difference of 0.6 meV per unit cell containing two formula units, between the in-plane and the out-of-plane magnetization directions; the energy difference (of the order of µeV per unit cell) between $x$ and $y$ magnetization directions is too small to determine the in-plane magnetic anisotropy [27]. Throughout this paper, we limit our discussion to the case with the magnetization $M$ along y, which has higher symmetry than with $M//x$. Change in the magnetization direction leads to different magnetic space groups corresponding to symmetry changes modifying the band structures exemplified by the different band crossings and anti-crossings seen in Fig. 1d.

Figures 1e, f show the calculated and measured (with photon energy $h\nu = 90$ eV) Fermi surface (FS) in the $k_x$-$k_y$ plane. The first Brillouin zone (BZ) boundary is marked by the orange hexagon. The electron pockets around the K points can be identified both from the calculated spectra and ARPES data. We conclude that the samples are mainly terminated by Sn by examining the Sn 4$d$ and Fe 3$p$ core-level spectra collected for photon energies 130 eV and 150

eV, as shown in Fig. 1g. The emission for the lower energy 130 eV photons is more surface sensitive, while it is more bulk sensitive for the 150 eV photons. The Sn $4d_{3/2}$ and $4d_{5/2}$ orbitals have an energy difference of 1 eV, yielding two peaks each of which splits into two further peaks attributable to the surface and bulk, with the lower binding energy associated with the bulk. The ratio between surface and bulk peaks varies with the escape depth of the electrons and therefore the outgoing electron energy and ingoing photon energy. The surface peaks are suppressed relative to those for the bulk when we increase $h\nu$ from 130 to 150 eV. For the Fe 3p levels, no splitting is observed and the spectra obtained with different photon energies hardly differ.

Fig. 2a displays a FS map measured in the $k_y$-$k_z$ plane. $k_z$-independent bands are observed around the *K*-point (marked by arrows), indicating a quasi-2D behavior. ARPES intensity plots at two $k_z$ planes are shown in Fig. 2b,c, respectively. The arrows are located at the same momentum as in Fig. 2a. As found in a recent ARPES study, a Dirac-like cone formed by quasi-2D Dirac fermions is present around the *K*-point [26]. In Fig. 2d, we plot the surface-projected DFT calculated band structure along $k_y$. At the *K*-point, we observe Dirac-like band dispersions as detected by ARPES.

For the UV photons used by ourselves and previous workers, the bands observed very near the Fermi energy appear two-dimensional. In particular, the FS map in the $k_y$-$k_z$ plane, shown in Figure 2a, consists of streaks parallel to the *z* axis, while the quasiparticle dispersions including Dirac-like cones, found in $E$-$k_y$ cuts (Figs. 2b and c) with different fixed $k_z$ do not differ noticeably, but are consistent with features shown in the surface-state calculations of Fig. 2d. To further examine the dimensionality of the bulk bands, we performed photon energy-dependent ARPES measurements using soft X-rays which probe beyond the surface layer. Figure 2e shows the resulting FS map in the $k_y$-$k_z$ plane. The strong features are no longer streaks along $k_z$, but rather lines which curve around the $k_z$ axis, as would be expected for a layered material with moderate coupling between layers. These lines are not related to the streaks seen for the UV photoemission (Figure 2a) because they are much closer to the 2D zone center, with $k_y \approx$ 0.25 rather than 0.5. Furthermore, dispersive features with clear periodicity, matching that of the extended Brillouin zone along $k_z$, are now observed in the $E$-$k_z$ cut displayed in Fig. 2f; what is striking is that these are remarkably Dirac-like, thus providing the first evidence for topological behaviour in the bulk of $Fe_3Sn_2$. The $E$-$k_y$ cuts in Fig. 2g and 2h show consistency with the bulk DFT calculations including a lower Dirac-like cone marked with dashed lines.

The lack of obvious closed orbits in Fig. 2e has implications for the orbital magnetoresistance (MR): when the magnetic field has a component perpendicular to the $k_y$-$k_z$ plane, we expect non-saturating MR, which agrees with transport measurements to be published separately. In addition, quantum oscillations should be much easier to find for fields along z than along the Kagome planes.

Having validated our DFT calculations via detailed comparison with ARPES, we turn now to their most remarkable outcome, namely discovery of Weyl nodes very close to the Fermi energy and their control via the magnetization direction. The nodes can exist in $\mathcal{P}$-invariant, ferromagnetic $Fe_3Sn_2$ by virtue of the crossing of majority and minority spin bands in addition to Weyl nodes between bands with the same spin polarization, which is the only option for paramagnetic Weyl semimetals in $\mathcal{P}$ symmetry broken systems. The Weyl nodes are dispersed differently for different magnetization directions because the directions control the system symmetry. Fig. 3 and Table I summarize the positions and energies of Weyl points within ±10 meV of $E_F$ for magnetization along the three major symmetry directions, which we each now consider in turn. First, if the magnetization is along $y$, the system is protected by space inversion symmetry $\mathcal{P}$, $C_{2y}$, and the mirror operation $\mathfrak{M}_y$. In the calculated band structures, eight pairs of Weyl nodes (four symmetry-inequivalent Weyl points) are found, as shown in Fig. 3a. Red and blue dots represent +1 and -1 chirality, respectively. The top view just shows the projections of Weyl points in the $k_x$-$k_y$ plane. $W_{y4}$ comprise of two pairs. In addition, we found the two-fold degenerate nodal-line structures on the $k_y$=0 plane (see Supplementary Figure 1) which are protected by $\mathfrak{M}_y$.

If the magnetization is along the $x$ direction, the system is protected only by space inversion symmetry. In this case, we can find ten pairs of Weyl points, all of which are symmetry inequivalent, as seen in Fig. 3b. Interestingly, even if $\mathfrak{M}_y$ is broken under $x$ magnetization, the distribution of Weyl points almost satisfies $\mathfrak{M}_y$, whereas the chirality doesn't respect the condition for the mirror operation. When the magnetization is along the $z$ direction, as above $T_f$, the system is protected by $\mathcal{P}$-symmetry, $C_{3,-z}$ and $C_{3z}$. In this case, within ±10 meV of $E_F$, six pairs of Weyl nodes (two symmetry-inequivalent Weyl points) exist, as shown in Fig. 3c.

Interesting new topological properties of $Fe_3Sn_2$ can be derived from level crossings between majority and minority spin bands as a function of momentum. Figure 4 shows the crossing associated with the Weyl point $W_{y1}$ closest to the Fermi level; the linear crossings are tilted and therefore we identify $W_{y1}$ as type-II. Without SOC (Fig. 4d), up and down spin channels are decoupled, and we see an unavoided level crossing. If we then polarize the spins along different directions with SOC turned on, the level crossing remains unavoided (M//y, Fig. 4a) or avoided (M//x and z, Figs. 4b and 4c).

In all panels of Fig. 4, the band dispersions are color-coded by the spin expectation values of the spin components along M from first-principles calculations with a full account of the one-center term of projector-augmented wave formalism. As the Weyl nodes are approached, these expectation values vanish, as one would expect near any magnetic level crossing where up and down spin wavefunctions are mixed in equal proportion. However, Figs 4e-4f reveal a new effect, namely the appearance of transverse quasiparticle polarization as the avoided Weyl node is approached. In other words, there is magnetic texture in k-space associated with the avoided Weyl nodes in this ferromagnet. Almost needless to say if a Weyl point were between the same spin channels the crossing would not affect the expectation value of the quasiparticle spin.

The Weyl nodes predicted by the DFT calculations have not been observed by ARPES measurements. One possible cause is the 3-fold rotational degeneracy of the magnetic domains in the samples. A typical beam size of synchrotron beam is around 100x100 $\mu m^2$, which is larger than the typical magnetic domain and therefore many magnetic domains with different band structures would be probed simultaneously using ARPES.

Our calculations and particularly the results in Figs. 3 and 4 also have other experimental consequences. First, the proximity of many Weyl nodes to the Fermi surface combined with their sensitivity to magnetization direction account for the so far mysterious spin reorientation transition, accompanied by dramatic evolutions in electrical properties, in $Fe_3Sn_2$ at the relatively low temperature of ~ 100 K: changes in the chemical potential which occur as the temperature changes will select between the different nearly degenerate electronic ground states associated with different magnetization directions.

Similar to temperature changes, very modest fields steering the magnetization can change the band structure, resulting in avoided level crossings. Further increases in fields can shift the states near the level crossings, but can also affect the non-trivial spin texture which we have discovered among the quasiparticles in momentum space. Therefore, magnetotransport measurements will contain multiple features related to a variety of phenomena, including Zeeman effects near Weyl nodes as well as unwinding of quasiparticle spin textures. Finally, considering that a magnetic field of a fraction of a Tesla is sufficient to change the magnetization direction in the plane and a field of the order of one Tesla is sufficient in $Fe_3Sn_2$ to alter the magnetization direction from in-plane to out of plane, the tunability by magnetic fields [28-29] can be exploited in future spin-based devices or devices based on Weyl fermions.

In summary, we have systematically explored the band structure of $Fe_3Sn_2$ using first-principles calculations and ARPES experiments. The calculations show a gapped massive Dirac-like spectrum, which agree with photoemission data that reveal a quasi-2D Dirac cone and 3D bulk electronic structure. The prediction of Weyl nodes and their dependence on magnetization orientation, together with the fact that $Fe_3Sn_2$ is a soft ferromagnet whose magnetization can be easily controlled by external magnetic field or modest temperature, opens opportunities for devices where the electronic structure is switched using either the chemical potential or magnetic field.

**Methods**

**Calculation methods.** Our theoretical calculations are based on density functional theory (DFT) [30-31] as implemented in the Vienna *ab initio* simulation package (VASP) [32-33] and the generalized gradient approximation of Perdew-Burke-Ernzerhof-type [34] is employed for the exchange-correlation energy. The electron-ion interaction is described by the projector augmented-wave method [35]. The *p* and *d* semi-core states are included in the Fe and Sn PAW datasets used, respectively. The simplified on-site Hubbard U correction [36] is added on Fe *d* orbitals and spin-orbit coupling is considered throughout all calculations. Wave functions are expanded in terms of plane waves with the kinetic energy cutoff of 450 eV and the ground-state charge density is evaluated on 24 x 24 x 24 *k*-point mesh. Weyl nodes are found using the *ab initio* tight-binding Hamiltonian in combination with the constrained version of simplex downhill method [37] starting from the regular *k* grid of 30x30x30. The *ab initio* tight-binding Hamiltonian is constructed using maximally-localized Wannier functions [38] as a basis from *ab initio* calculations. This Hamiltonian is also used in calculating momentum-resolved density of states using the iterative Green's function method [39]. The chirality of Weyl points is evaluated by using the Bloch states from *ab initio* calculations. For this purpose, we introduced the geodesic polyhedron enclosing each Weyl point which consists of triangles and quadrilaterals, and then used the Fukui method [40] for this geodesic polyhedron. For all Weyl points in our study, the 21×20 partitions were enough to obtain the converged value of chirality.

**Growth of $Fe_3Sn_2$ crystals**

High quality $Fe_3Sn_2$ single crystals were grown by chemical vapor transport method. Firstly, the polycrystalline precursors were synthesized by a solid-state reaction using stoichiometric iron (Alfa Aesar, ＞99.9%) and tin (Alfa Aesar, ＞99.9%) in sealed and evacuated quartz tubes. The mixtures were heated up to 800 ℃ and maintained at this temperature for 7 days before quenching in water. The sintered $Fe_3Sn_2$ was thoroughly grinded and sealed with $I_2$ (~ 4 mg/cm$^3$) in a quartz tube under vacuum, and kept in a temperature gradient 720 ℃ to 650 ℃ for two weeks. $Fe_3Sn_2$ single crystals were obtained at the cold end. The laboratory X-ray diffraction measurements, which had been done at room temperature using Cu $K_α$ radiation on Rigaku TTR3 diffractometer have proven that the obtained crystals are single phase with rhombohedral structure of space group $R\bar{3}m$.

**Angle-resolved photoemission spectroscopy.** Clean surfaces for ARPES measurements were obtained by cleaving $Fe_3Sn_2$ samples in situ in a vacuum better than 5x10$^{-11}$ Torr. ARPES measurements were performed in the Swiss Light Source at the SIS-HRPES beamline with a Scienta R4000 analyzer, and ADRESS beamline with a SPECS analyzer. At the SIS-HRPES beamline, the energy and angular resolutions were set to 15 ~ 30 meV and 0.2°, respectively. At the ADRESS beamline, the data were collected with an overall energy resolution of the order of 50 ~ 80 meV for FS mapping and 40 ~ 60 meV for high-resolution cut measurements. All data were taken at 15 K.

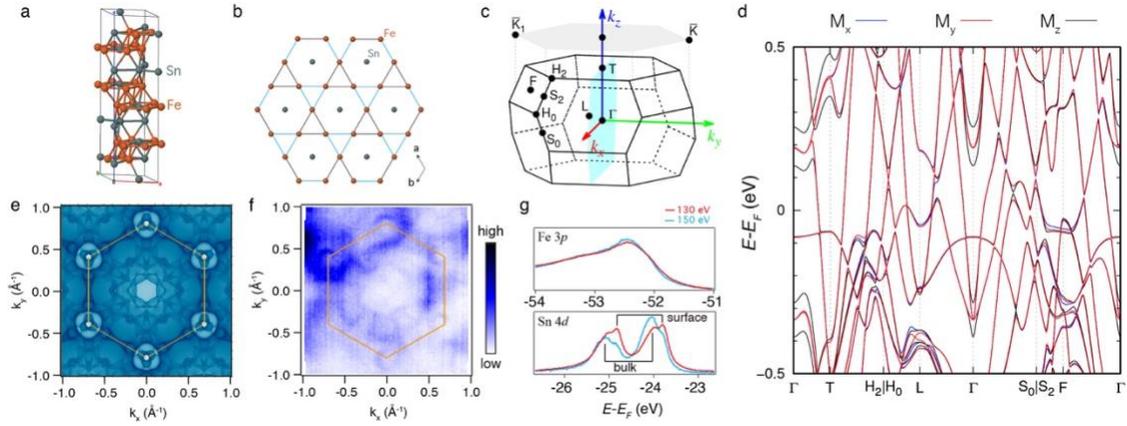

Figure 1 (a) Crystal structure and (b) the kagome structure of $Fe_3Sn_2$. The black and blue lines in (b) indicate the longer and shorter equilateral triangles, respectively. (c) Bulk Brillouin zone (BZ) and the gray-shaded (111) surface BZ of $Fe_3Sn_2$. High-symmetry points are indicated by black dots with the corresponding labels. The cyan-shaded plane represents the mirror plane in the case of $y$ magnetization. (d) Bulk band structure of $Fe_3Sn_2$ along several high-symmetry lines from first-principles DFT+U calculations with the inclusion of SOC. Band structures for different magnetization directions are indicated by different colors. (e) Calculated FS map in the $k_x$-$k_y$ plane and the corresponding FS map taken with $h\nu = 90$ eV (f). (g) Fe $3p$ (upper panel) and Sn $4d$ (lower panel) core-level spectra taken with $h\nu = 130$ and $150$ eV.

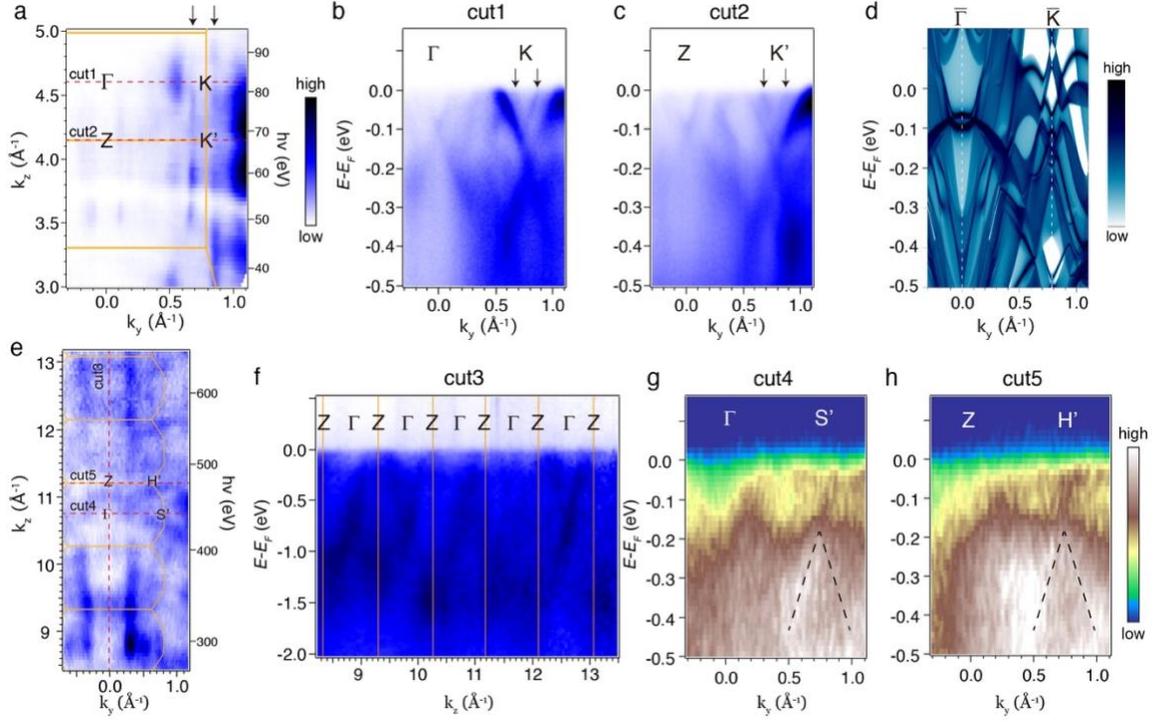

Figure 2 Electronic structure of $Fe_3Sn_2$. (a) FS map in the $k_y$-$k_z$ plane at $k_x = 0$, acquired with UV light. Orange solid lines indicate Brillouin zones. (b, c) ARPES intensity plot at different $k_z$ planes, as marked with red dashed lines in (a), respectively. (d) Calculated momentum-resolved density of states for Sn-terminated surface along the $k_y$ direction through $\bar{\Gamma}$ and $\bar{K}$ Gamma points. (e) FS map in the $k_y$-$k_z$ plane at $k_x = 0$, acquired with soft X-ray. (f) ARPES intensity plot obtained at $k_y = 0$. (g-h) ARPES intensity plot of cut4 and cut5 indicated in (e).

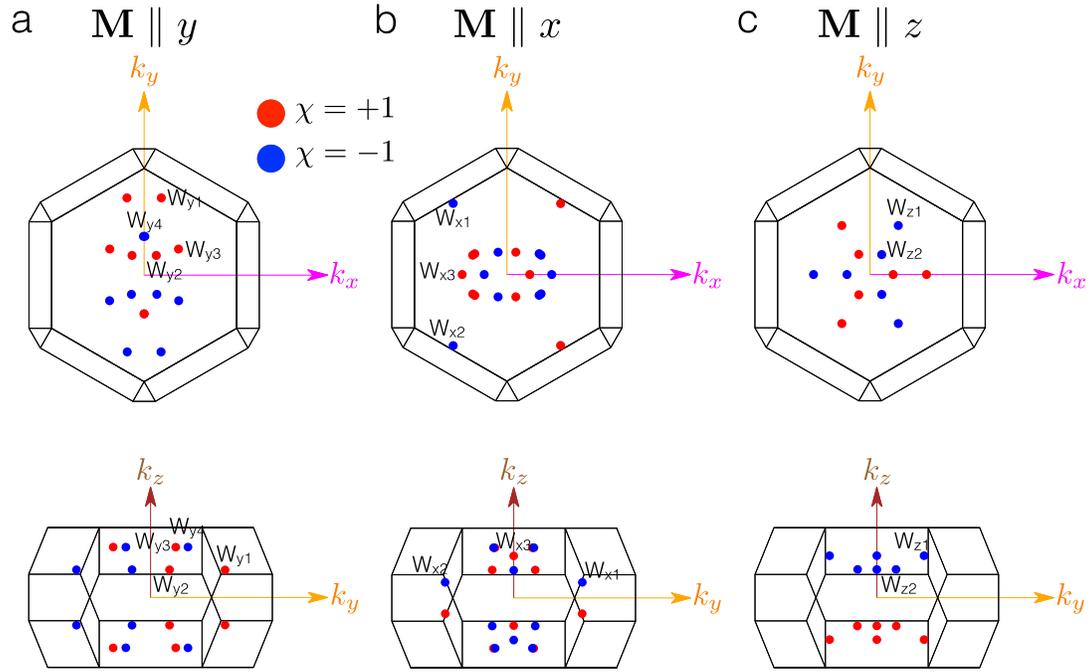

Figure 3 Distribution of Weyl points from *ab initio* calculations for (a) *y* magnetization, (b) *x* magnetization, and (c) *z* magnetization. Here, we show only Weyl points with their energies within ±10 meV with respect to the Fermi energy. In the case of *x* magnetization, only three symmetry-inequivalent Weyl points are labeled.

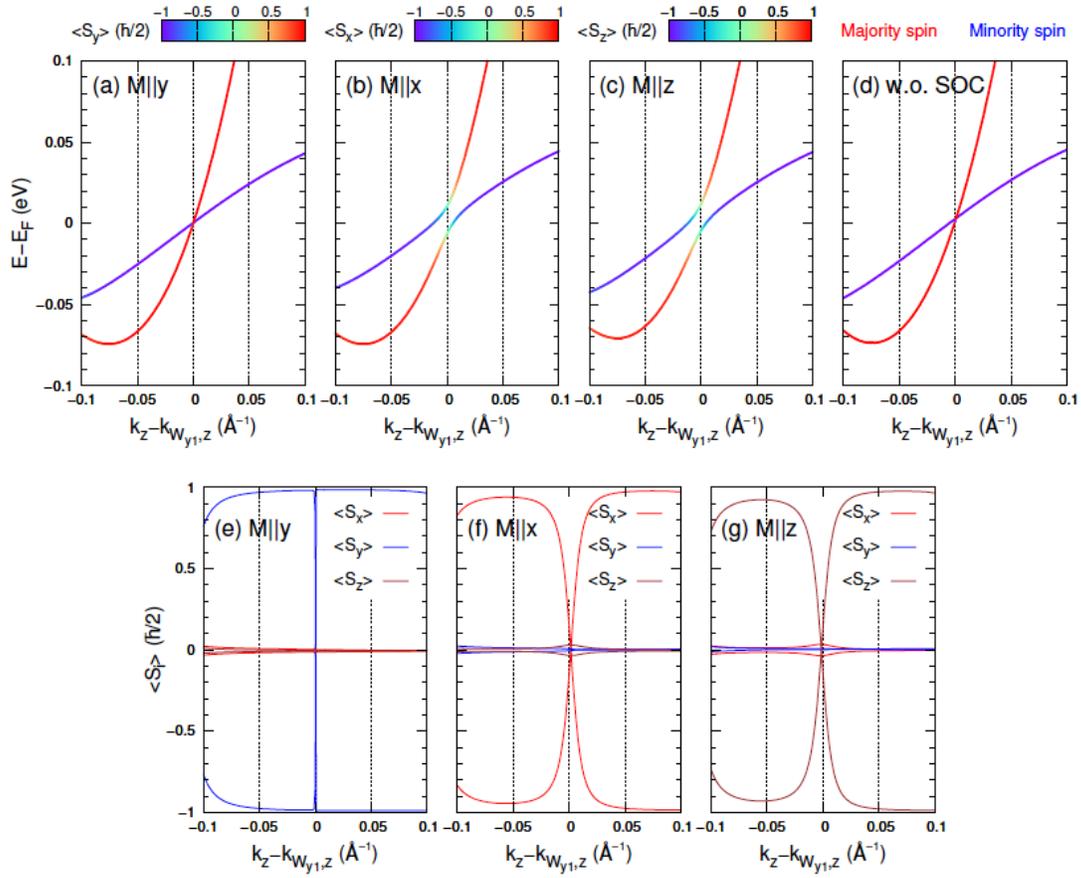

Figure 4: Band dispersions around the Weyl point ($W_{y1}$) closest to the Fermi level (a,b,c) with SOC and (d) without SOC. In the case with SOC, (a) refers to the *y* magnetization, (b) the *x* magnetization, and (c) the *z* magnetization. Bands are color-coded by the spin expectation value of the corresponding major spin component. (e-g) The evolution of spin expectation values for (e) *y*, (f) *x*, and (g) *z* magnetizations.

| $y$ mag. | $k_x$ | $k_y$ | $k_z$ | $E-E_F$ | chirality |
|---|---|---|---|---|---|
| $W_{y1}$ | 0.113 | 0.506 | -0.187 | 0 | +1 |
| $W_{y2}$ | 0.079 | 0.129 | 0.189 | -6 | +1 |
| $W_{y3}$ | 0.225 | 0.170 | -0.342 | -1 | +1 |
| $W_{y4}$ | 0.004 | 0.254 | -0.343 | 4 | -1 |
| $z$ mag. | | | | | |
| $W_{z1}$ | 0.185 | 0.321 | 0.286 | -9 | -1 |
| $W_{z2}$ | 0.077 | 0.131 | 0.192 | -4 | -1 |
| $x$ mag. | | | | | |
| $W_{x1}$ | -0.352 | -0.465 | 0.106 | -7 | -1 |
| $W_{x2}$ | -0.352 | 0.465 | 0.106 | -7 | -1 |
| $W_{x3}$ | -0.293 | 0 | 0.286 | -6 | +1 |
| $W_{x4}$ | 0.058 | -0.146 | 0.191 | -6 | +1 |
| $W_{x5}$ | 0.058 | 0.146 | 0.191 | -6 | +1 |
| $W_{x6}$ | 0.149 | 0 | -0.188 | -6 | +1 |
| $W_{x7}$ | 0.227 | 0.122 | -0.341 | 8 | -1 |
| $W_{x8}$ | 0.227 | -0.122 | -0.341 | 8 | -1 |
| $W_{x9}$ | 0.211 | -0.136 | 0.342 | 9 | -1 |
| $W_{x10}$ | 0.211 | 0.137 | 0.342 | 9 | -1 |

Table 1: The positions and energies of Weyl points within ±10 meV with respect to the Fermi Energy ($k$ in 1/Ang, $E$ in meV units). These Weyl points were found from the *ab initio* tight-binding Hamiltonian and these locations might be different from those in full *ab initio* calculations. However, we guarantee all Weyl points reside within the distance of 0.005 1/Ang from the positions above by evaluating their chiralities in a fully *ab initio* way.

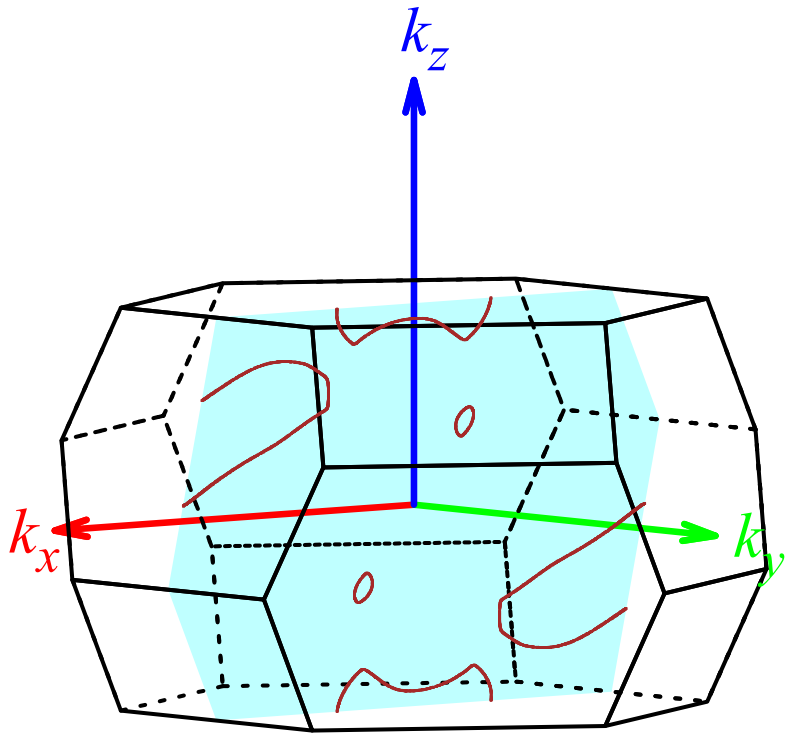

Supplementary Figure 1: Nodal lines resulting from band crossings between *N*-th and (*N*+1)-th bands (*N*: # of electrons in the unit cell) for the *y* magnetization.

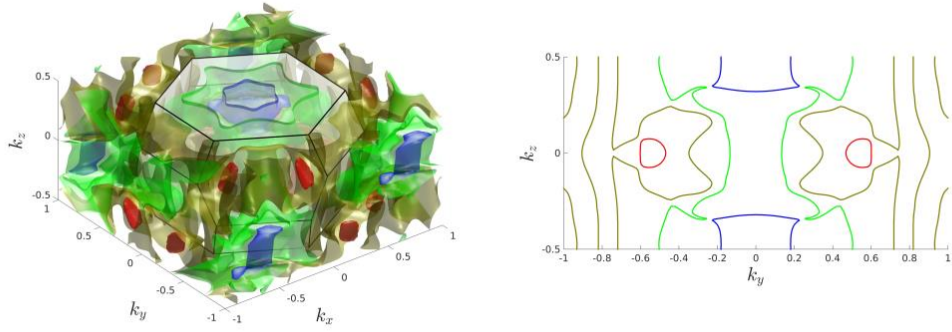

Supplementary Figure 2: FS mapping for the *y* magnetization (left) and cross section of FS mapping in the $k_z$-$k_y$ plane showing open orbits (right).